\documentclass[pre,twocolumn]{revtex4}
\usepackage[english]{babel}
\usepackage{amsmath}
\usepackage{amssymb}
\usepackage{epsfig}

\begin{document}
\title{Formation of nonlinear waves in decelerated centrifuges of noncircular cross-section}
\author{Victor P. Ruban}
\email{ruban@itp.ac.ru}
\affiliation{Landau Institute for Theoretical Physics RAS,
Chernogolovka, Moscow region, 142432 Russia}

\date{\today}

\begin{abstract}
Planar flows with a free boundary in a partially filled and nonuniformly rotating
container, with a strongly noncircular shape of the cross-section, are investigated 
numerically within the ideal fluid approximation. Vorticity is assumed constant
across the fluid, thus allowing us to apply the recently developed, highly
efficient numerical method based upon exact equations of motion of the free
boundary in terms of conformal variables and on the fast Fourier transform algorithms.
It is shown that decelerated rotation of such centrifuge leads to formation of
strongly nonlinear breaking waves with sharp crests, and the wave overturning 
occurs either in the rotation direction or against it, depending on value of the 
(negative) angular acceleration.
\end{abstract}

\maketitle

Nonstationary flows with a free boundary have been an important part of
the classical hydrodynamics, mainly due to the inexhaustible interest to
waves on the sea surface and on other bodies of water. Gravity water waves are
excitations caused by the presence of the uniform gravity field and propagating
in the two horizontal directions. The amount of published scientific articles
on this subject is indeed huge. Essentially, in most theoretical models
the approximation of an ideal incompressible fluid is employed, together with
the potentiality condition for the velocity field. Cartesian coordinates are
quite natural in description of small deviations from the equilibrium plane,
so below we will conventionally call such waves ``Cartesian waves''.

\begin{figure}
\begin{center}
\epsfig{file=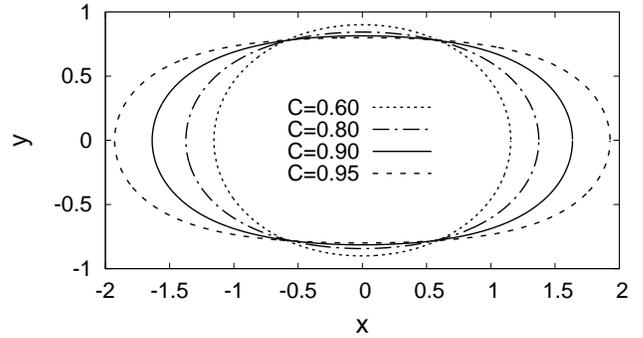, width=88mm}
\end{center}
\caption{Cross-sections of the centrifuge as given by Eq.(\ref{cross-section}),
for several values of $C$.
}
\label{CS} 
\end{figure}

In contrast to ``Cartesian'' waves, surface waves in partially filled, 
rapidly rotating containers --- centrifuges --- have been studied so far 
in a considerably less extent (see, e.g., \cite{Phillips,IKC2004,IKP2005},
and references therein). It should be said that in a fast rotating coordinate
system, the efficient gravity field is created by the centrifugal force
in the radial direction from the rotation axis, and this field is dominating
over the usual gravitational acceleration $g$. Therefore, the angular
coordinate plays here the role analogous to the role of one of the horizontal
coordinates for ``Cartesian'' waves, whereas the radial coordinate is analogous
to the vertical coordinate in the uniform gravity field. Besides that, a very
essential influence on the dynamics of surface waves in centrifuges is caused
by the Coriolis force. The problem is especially difficult when the shape of
the container cross-section is different from circular, and when the rotation
occurs with an angular acceleration. The fluid dynamics in such systems is
interesting both for the fundamental science and for possible technological
applications, and therefore it is deserving of serious attention. In this work,
we consider decelerated centrifuges with cross-sections as shown in Fig.\ref{CS},
thus making a new step in theoretical investigation of nonlinear water waves.

Let us start with simplifying approximations. If the container of a typical
transverse size $L$ is rotating for a long time with a constant angular velocity
$\Omega$ around some horizontal axis, then (under condition $g/(L\Omega^2)\ll 1$)
an almost rigid-body regime of rotation arises, and all the dissipative vortex
structures practically disappear. After that, an applied external action, for
example linear acceleration of the container axis or its rotational acceleration
(in the case of noncircular shape), leads to deformation and subsequent
nontrivial dynamics of the free internal surface. It is important that with
sufficiently large values of $\Omega$, inertia is dominating over dissipation
for some initial time period, until large gradients have been formed near
the free surface, or until boundary-layer separation has occurred near possible
wall irregularities. On this initial stage, the ideal-fluid approximation
is quite admissible, and it is able to provide a rich information about 
wave dynamics in centrifuges.

The second simplification is to consider the problem in the class of planar flows.
The two-dimensionality makes possible application of time-dependent conformal
mappings (of the flow domain) and the mathematical apparatus of the corresponding
analytical functions \cite{Ovs1,Ovs2}. This approach allows one to reduce the
spatial dimensionality of the dynamical system to 1D (the evolutional system of
equations contains only one independent spatial variable being a parameter along
the free boundary). The equations themselves can be written in an exact form
suitable for numerical simulations using fast Fourier transform routines
\cite{Z1,Z2,Z3,D1}. At present, equations of motion of the free surface in
conformal variables have been a working tool for many researchers (see Refs.
\cite{Z4,Z5,Z6,Z7,ChoiC1999,LHChoi2004,Sh2006,Sh2008,Sh2010,ZSh2010,ZSh2012,
ChS2005,Ch2009,Sl2009,Choi2009,LDK2013,ZK2014,L2016,TB2016,T2016,BP2016,LDS2017,
DLZ2019,DDLZ2019,D2019,GDVW2019,FMN2019,KDG2019}, and many references therein).
Mainly, waves over infinitely deep water are simulated, but the method 
was also generalized to the case of a curved static bottom, by introducing
a composition of two conformal mappings \cite{R1}. The second, auxiliary
mapping is realized by some given analytical function, and it corresponds to
the bed profile. Waves over a time-dependent bottom were then considered in
Ref. \cite{R2}, and shear flows with a free boundary over variable depth were
modeled for the first time by the composite conformal mapping method in
\cite{R3}. High efficiency and practical convenience of this method were
confirmed in solution of some meaningful physical problems \cite{R4,R5,R6,R7,R8}.
Finally, for uniformly swirled flows in centrifuges, conformal variables were
introduced and approved in the recent work \cite{R2020}.

In contrast to typical ``Cartesian'' waves propagating over initially static 
body of fluid, for consideration of planar flows in centrifuges, instead of the 
potentiality condition, it is more natural to put the condition of constant
vorticity (in the laboratory frame of reference). Therefore in a uniformly 
rotating coordinate system the velocity field is purely potential:
${\bf V}=(V^{(x)},V^{(y)})=(\varphi_x, \varphi_y)=(\theta_y,-\theta_x)$, 
where the subscripts denote partial derivatives, function $\varphi(x,y,t)$
is the velocity potential, and $\theta(x,y,t)$ is the corresponding
harmonically conjugate stream function. It is very important that
$\varphi(x,y,t)$ obeys the generalized Bernoulli equation \cite{R2020}, 
\begin{eqnarray}
\varphi_t+(\varphi_x^2+\varphi_y^2)/2-\Omega^2(x^2+y^2)/2+2\Omega\theta 
+&&\nonumber\\
+g(t)(y\cos\Omega t+x\sin\Omega t)+\tilde P=0,&&
\label{Bernoulli_generalized}
\end{eqnarray}
where $\tilde P$ is the pressure divided by (constant) fluid density.
The effective vertical gravity field $g(t)$ depends on time if the rotation
axis experiences vertical accelerations. Besides the potential of centrifugal
force, the term $2\Omega\theta$ is present here, which is due to the Coriolis
force. Equation (\ref{Bernoulli_generalized}), taken on the free internal boundary,
determines the wave dynamics in centrifuge (together with two kinematic conditions,
on the free boundary and on the container wall). When expressed in terms of the
conformal variables, the equations serve as the basis for a highly accurate
numerical method, described in detail in work \cite{R2020}. It is very essential
that employment of composite conformal mappings allows one to deal easily with
centrifuges of noncircular shape, if the wall profile is given parametrically
as an image of the unit circle under some conformal mapping of its interior,
i.e.,  $X+iY=F(e^{i\zeta})$, $0\leq\zeta\leq 2\pi$, with an analytical function
$F(W)$. 

\begin{figure}
\begin{center}
\epsfig{file=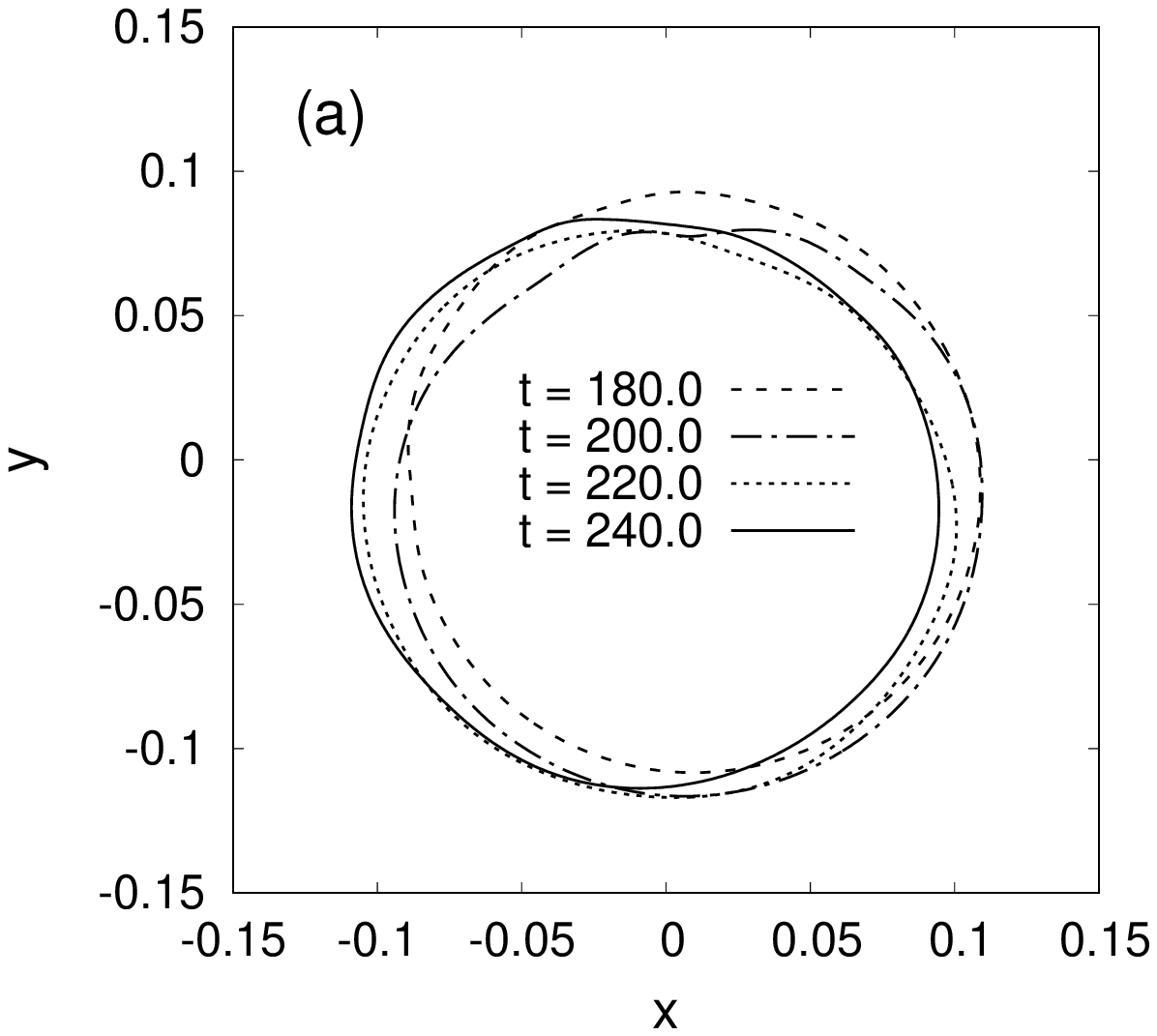, width=73mm}\\
\epsfig{file=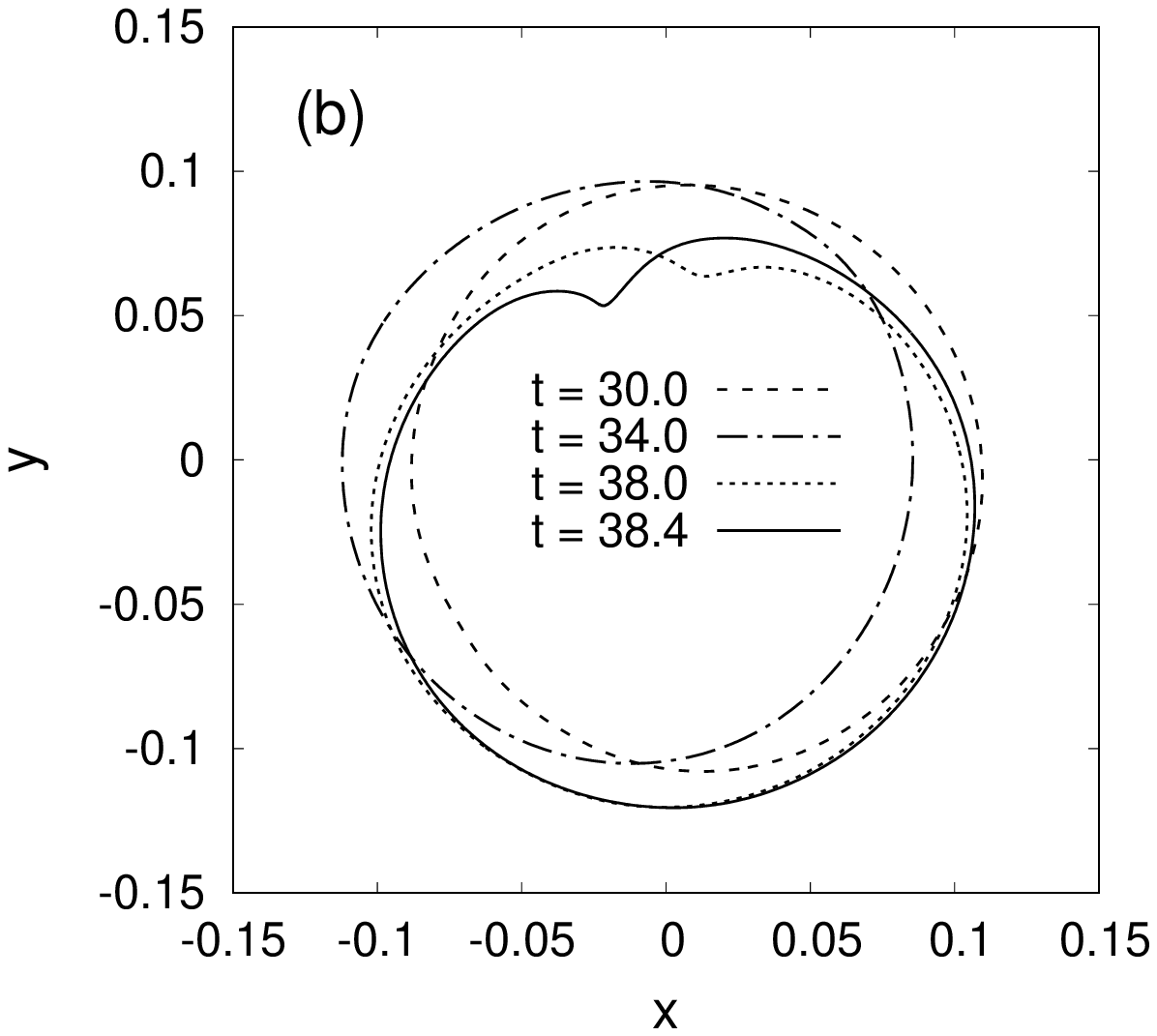, width=73mm}\\
\epsfig{file=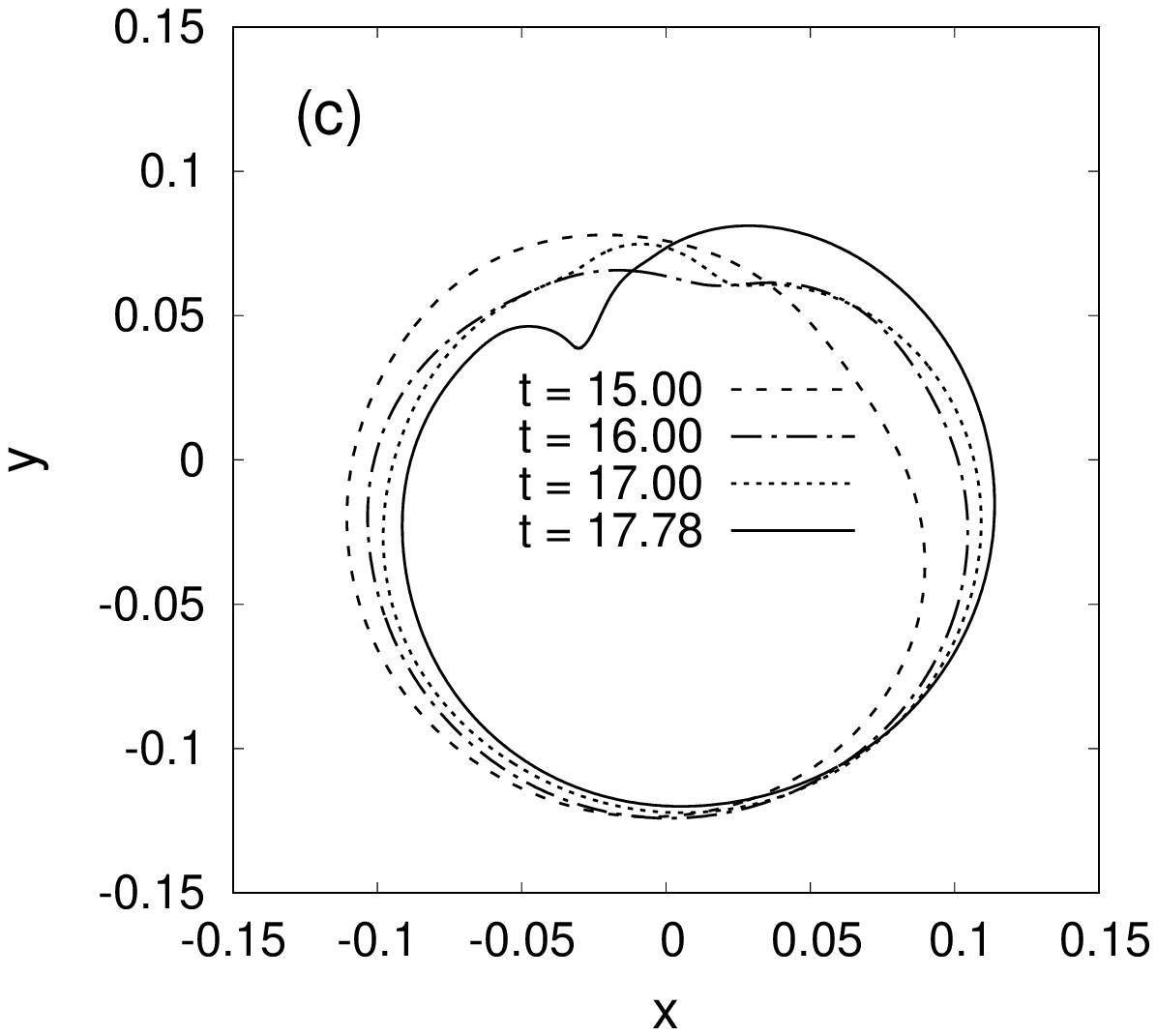, width=73mm}
\end{center}
\caption{Numerical example of the influence of parameter $g$ upon the time
when nonlinear waves rise in a uniformly rotating ($\alpha=0$) centrifuge:
(a) $g=0.018$, (b) $g=0.021$, (c) $g=0.025$.
The remaining parameters are $C=0.9$, $s_0=0.1$.
}
\label{g-1-2-3} 
\end{figure}

In this work, the method is applied to simulate strongly nonlinear processes
taking place in centrifuges with elongated shape of the cross-section under their
angular deceleration, when (non-dimensionalized by $\Omega$) angular velocity
varies with time according to the linear law $\omega(t)=1-\alpha t$. Such a
statement of the problem seems rather natural. For the centrifuge cross-section
shape (at the time moment $t=0$), the following function was chosen,
\begin{equation}
X(\zeta)+iY(\zeta)=\frac{1}{2C}\ln\frac{1+Ce^{i\zeta}}{1-Ce^{i\zeta}},
\label{cross-section}
\end{equation}
with parameter $0<C<1$. All the lengths are non-dimensionalized here to $L$.
As $C\to 0$, we have the unit circle here, whereas in the limit $C\to 1$
the domain elongates to the horizontal stripe of width $\pi/2$. In Fig.\ref{CS},
examples of cross-sections for several intermediate values of $C$ are presented.
At an arbitrary time moment, expression (\ref{cross-section}) is multiplied by
$\exp(i\int_0^t \omega(t) dt)$. It is clear in advance that in the coordinate
system attached to the container of such elongated shape, under deceleration
of its rotation, a shear component of the flow appears, which interacts with
the free boundary and tends to deform it. A shear flow around the free cavity
is accompanied by formation of surface waves. It is quite obvious that
investigation of nonlinear fluid dynamics in such geometry is impossible by the
present-day-existing analytical methods. Meanwhile, the numerical method copes
with the problem easily, keeping a high accuracy typically exceeding 8-10
decimal places.

In the performed numerical experiments, initial shape of the free boundary
was given parametrically as $x(u)+iy(u)=F(s_0 e^{i u})$, with some parameter
$s_0$. Under condition $s_0C\lesssim 0.4$, it was close to the circle of radius
$s_0$. The initial velocity potential was equal to zero.

It should be said that in the coordinate system attached to the container,
the vertical gravity field plays the role of a time-dependent driving force
capable to excite waves. Therefore it was also taken into account, and its
non-dimensionalized to $L\Omega^2$ value was chosen as $g=0.02$ in the main
series of our numerical simulations. Separately it was checked that under uniform
rotation (when $\alpha=0$), even an essentially larger value as $g=0.06$ does not
lead to formation of nonlinear waves for a rather long time period, at least
for values $s_0 \sim 0.4$ that are not too large and not too small. In general,
for each $s_0$ there exists a characteristic value $g_*(s_0)$, near which the
time of nonlinear wave formation changes abruptly. The function $g_*(s_0)$
turns out to be increasing. For example, with $s_0=0.1$ (smaller cavity size),
the transient regime takes place just near $g=0.02$, as it follows from 
Fig.\ref{g-1-2-3} where profiles of the free boundary at different time moments
are shown for three numerical experiments (the rotation occurs anti-clockwise).
It is seen that at $g>0.02$ the nonlinear wave rises just after a few revolutions
of the centrifuge, whereas at $g=0.018$ big waves are still absent after dozens
of revolutions. Since at small $s_0$ the free surface is in the deep-water
regime, the relevant parameter is the ratio  $g/s_0$. From here we obtain
a rough empirical estimate $g_*\approx 0.2 s_0$. Such a tendency is explained by
weakening of the stabilizing effect of centrifugal acceleration at smaller
distances from the rotation axis.

\begin{figure}
\begin{center}
\epsfig{file=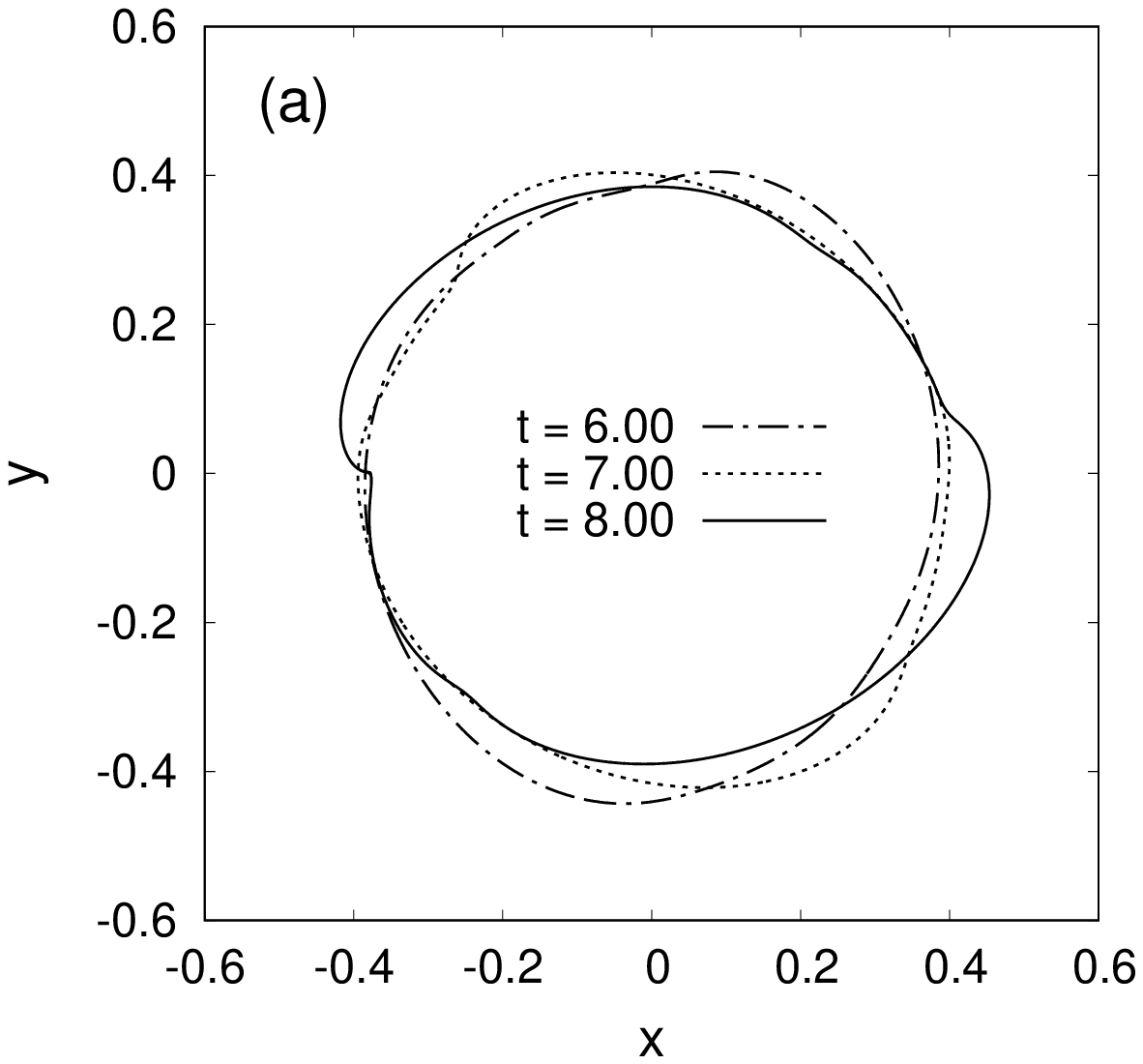, width=73mm}\\
\epsfig{file=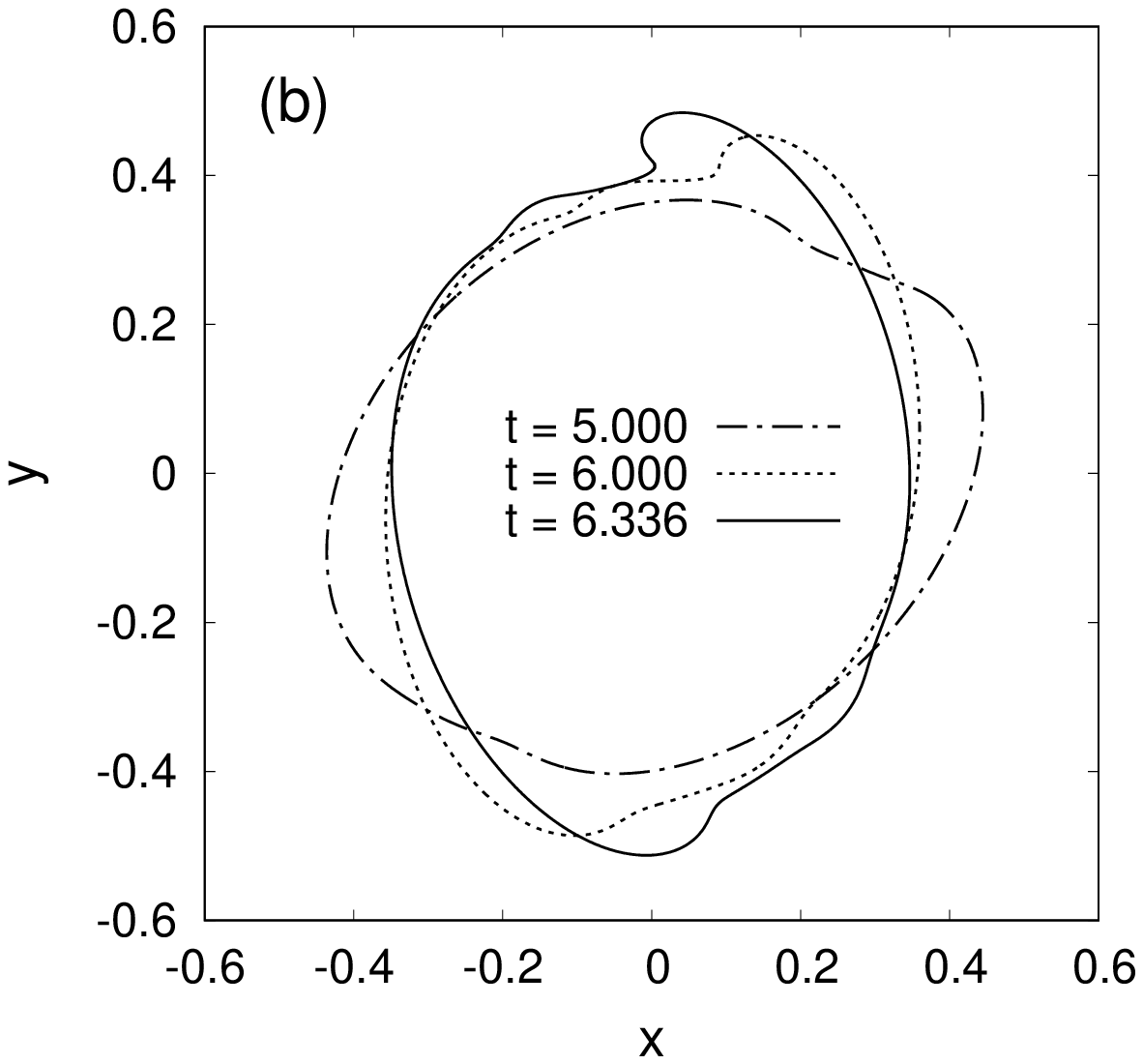, width=73mm}\\
\epsfig{file=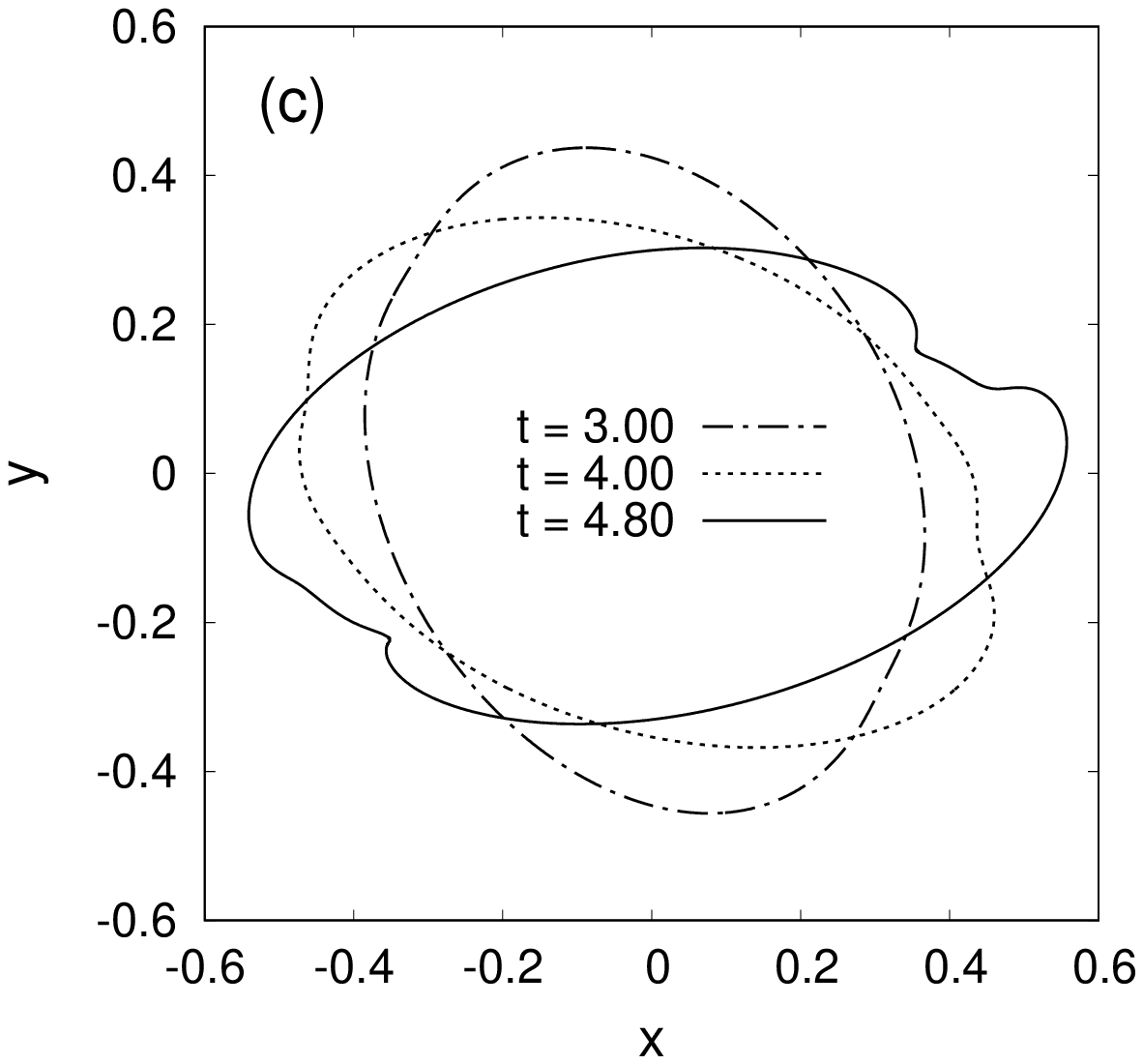, width=73mm}
\end{center}
\caption{Nonlinear waves with sharp crests on the free fluid boundary,
formed at different angular decelerations of the centrifuge:
(a) $\alpha=0.02$, (b) $\alpha=0.04$, (c) $\alpha=0.08$.
The remaining parameters are $g=0.02$, $C=0.9$, $s_0=0.4$. 
}
\label{alpha-1-2-3} 
\end{figure}

\begin{figure}
\begin{center}
\epsfig{file=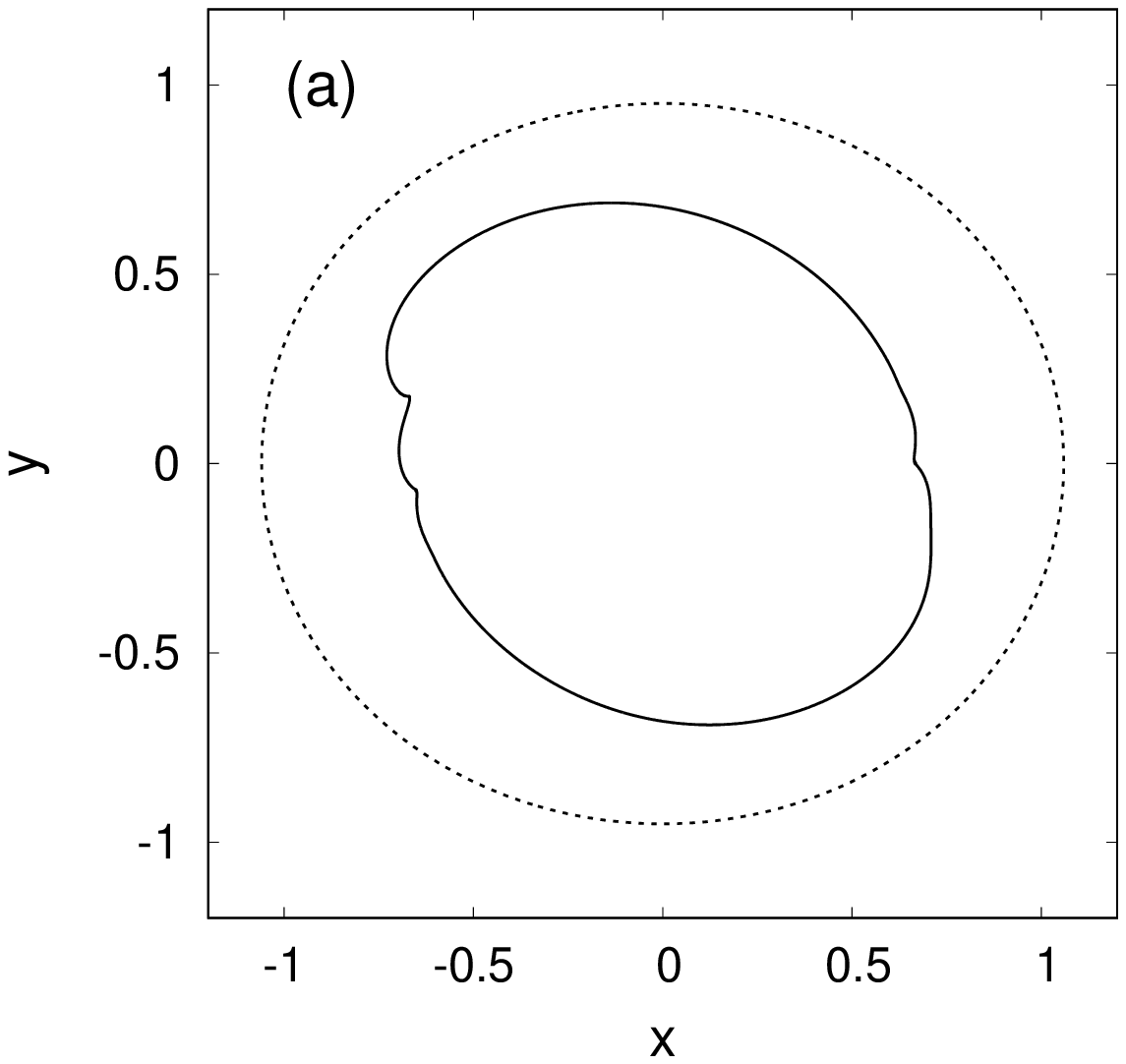, width=75mm}\\
\epsfig{file=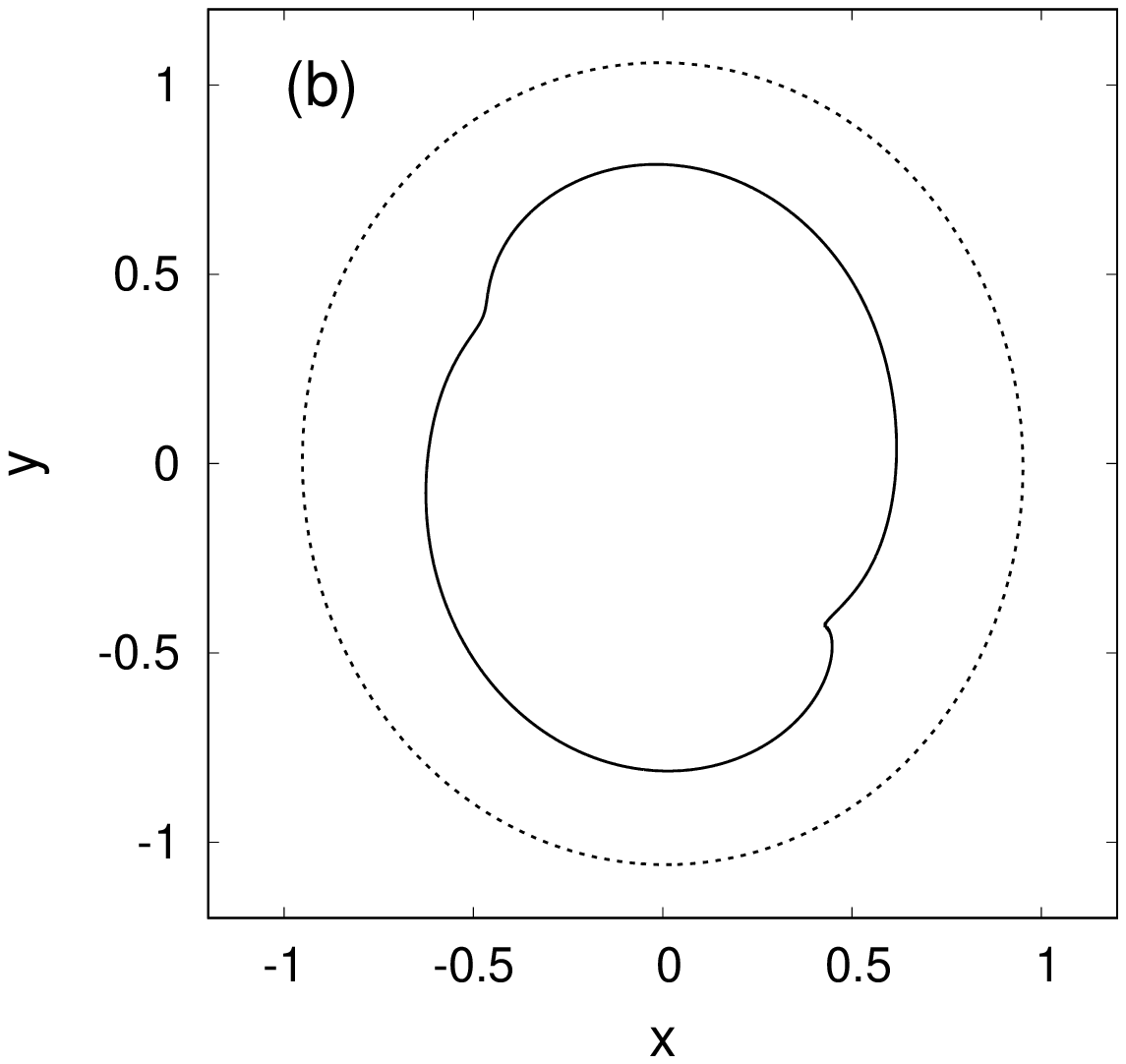, width=75mm}
\end{center}
\caption{Nonlinear waves at $g=0.02$, $C=0.4$, $s_0=0.7$:
(a) $\alpha=0.02$, $t=19.52$; (b) $\alpha=0.10$, $t=12.24$.
The solid line is for the free boundary of the fluid, and
the dotted line is for the wall of the centrifuge.
Note that in both examples the sharp crests overturn clockwise.
}
\label{C04} 
\end{figure}

Finally, let us pass to the main results for decelerated centrifuges. The simulations
have shown that arising shear flow leads the shape of free boundary away from
being circular and ``skews'' it, and after that several strongly nonlinear waves 
with steep and sharp crests appear in the system. Examples of such behavior are
presented in Fig.\ref{alpha-1-2-3} for $C=0.9$, $s_0=0.4$. We emphasize that, in
contrast to Fig.\ref{g-1-2-3} where the waves appeared always in the upper sector
of the free cavity due to the action of gravity force, here we clearly observe 
the tendency to wave emergence in two diametrically opposite directions, which
fact is related to the corresponding symmetry of the centrifuge cross-section.
The gravity force however slightly breaks the wave symmetry. It is also seen that
overturning of the wave crests occurs against the direction of centrifuge rotation
at small values of the angular deceleration $\alpha$, whereas with increase of this
parameter the crests overturn already in the rotation direction. The reason for
this is not clarified yet. 

Qualitatively similar results were obtained also for somewhat smaller cavity sizes,
down to $s_0\approx 0.15$, but a characteristic value $\alpha_*$, at which
the overturning direction is changed, slightly decreases there (from 
$\alpha_*\approx 0.06$ at $s_0=0.4$ to $\alpha_*\approx 0.04$ at $s_0=0.15$).
Besides that, at smaller $s_0$ the diametrical asymmetry increases.

With the purpose of comparison, simulations for a less filled ($s_0=0.7$) and not so
deformed ($C=0.4$) centrifuge were also carried out. Of course, with such geometrical
parameters, one cannot already speak about a shear flow around the cavity, but rather
about immediate wave interaction with an azimuthally nonuniform bottom profile, 
on the general background of accelerated angular stream. Nonlinear waves with
sharp crests appear here as well. The corresponding examples are given in
Fig.\ref{C04}.

It is clear that with the formation of a sharp crest, the stage of applicability
of the ideal fluid approximation comes to the end. Correspondingly, next stages 
require completely different approaches. Likely, the further evolution in real
conditions includes appearance of ``white caps'' and three-dimensional turbulent
domains. The number of involved degrees of freedom rapidly increases by orders of
magnitude (depending upon Reynolds number), and it can make subsequent modeling
quite problematic, sometimes even practically unavailable. That circumstance,
however, cannot devaluate the importance of investigation of the first,
inviscid stage. 

To conclude, in this work an initial stage of the dynamics of free fluid surface
in decelerated centrifuges with noncircular shapes has been accurately modeled
for the first time. The perfect fluid approximation, together with conformal
mappings, allowed us to observe evolution of nonlinear waves up to the breaking
of their crests. As one of the practical conclusions, we can say that in the
deep-water regime the main dissipative structures in a nonuniformly rotating
centrifuge will first appear most probably near the free surface, rather than
near a smooth wall.

As in every profound problem, here new results generate many new questions too. 
The subsequent work on the problem can be aimed, in particular, on the search
of new interesting numerical solutions in different parametric domains, and on the
achievement of qualitative understanding of the basic properties of observed
nonlinear wave structures.

\end{document}